\documentclass[11pt]{article}
\usepackage[a4paper, top=2.5cm, bottom=2.5cm, left=2cm, right=2cm]{geometry}

\usepackage{amsmath}
\usepackage{amsfonts}
\usepackage{amssymb}
\usepackage{amsthm}
\usepackage{stmaryrd} 
\usepackage{faktor}
\usepackage{braket}
\usepackage{empheq}
\usepackage{graphicx}
\usepackage{subfigure}
\usepackage{hyperref}
\usepackage{footmisc}
\usepackage{appendix}
\usepackage{epsf}
\usepackage{epsfig}
\usepackage{tikz,tikz-3dplot}
\usepackage{tikz-cd}
\usetikzlibrary{decorations.markings,decorations.pathmorphing}
\usepackage{caption}
\usepackage{pgfplots}
\usepackage[utf8]{inputenc}
\usepackage[english]{babel}
\usepackage[T1]{fontenc}
\usepackage{sectsty}
\usepackage{authblk}

\usepackage{tikz}
\usetikzlibrary{matrix,positioning,fit}

\usepackage[style=numeric-comp,sorting=none]{biblatex}
\subsubsectionfont{\normalfont\itshape}

\renewcommand{\bar}{\overline}
\renewcommand{\leq}{\leqslant}
\renewcommand{\geq}{\geqslant}

\newcommand{\bbone}{{\text{\usefont{U}{bbold}{m}{n}\char49}}}

\renewcommand{\H}{{\mathcal{H}}\xspace}
\renewcommand{\L}{{\mathcal{L}}\xspace}

\newcommand{\A}{{\mathcal{A}}\xspace}

\DeclareMathOperator{\tr}{tr}
\DeclareMathOperator{\stab}{Stab}

\theoremstyle{plain}
\newtheorem{theorem}{Theorem}[section]
\newtheorem{proposition}[theorem]{Proposition}
\newtheorem{corollary}[theorem]{Corollary}
\newtheorem{lemma}[theorem]{Lemma}

\theoremstyle{definition}
\newtheorem{definition}[theorem]{Definition}
\newtheorem{remark}[theorem]{Remark}
\newtheorem{example}[theorem]{Example}

\usepackage{xcolor}
\usepackage{xspace}
\usepackage{ifthen}
\newcommand{\showcomments}{true}

\newcommand{\choups}[1]%
{\ifthenelse{\equal{\showcomments}{true}}%
{{\color{violet}{\small \textbf{C:} #1}}}{\xspace}}%

\newcommand{\gui}[1]%
{\ifthenelse{\equal{\showcomments}{true}}%
{{\color{purple}{\small \textbf{G:} (#1)}}}{\xspace}}%

\newcommand{\andrea}[1]%
{\ifthenelse{\equal{\showcomments}{true}}%
{{\color{orange}{\small \textbf{A:} #1}}}{\xspace}}%

\title{On the emergence of preferred structures\\in quantum theory}
\date{}

\author[1,2]{Antoine Soulas\thanks{\texttt{antoine.soulas@univie.ac.at}}}
\author[3,4,5]{Guilherme Franzmann\thanks{\texttt{guilherme.franzmann@su.se}}}
\author[2,3]{Andrea Di Biagio\thanks{\texttt{andrea.dibiagio@oeaw.ac.at}}\vspace{0.2cm}}

\affil[1]{Quantum Optics, Quantum Nanophysics and Quantum Information, Faculty of Physics, University of Vienna, Vienna, Austria \vspace{0.2cm}} 

\affil[2]{Institute for Quantum Optics and Quantum Information (IQOQI), Austrian Academy of Sciences, Vienna, Austria \vspace{0.2cm}}

\affil[3]{Basic Research Community for Physics e.V., Mariannenstraße 89, Leipzig, Germany \vspace{0.2cm}}

\affil[4]{Nordic Institute for Theoretical Physics (NORDITA) and Department of Physics, Stockholm University, Stockholm, Sweden \vspace{0.2cm}}

\affil[5]{Department of Philosophy, Stockholm University, Stockholm, Sweden}



\begin{document}

\maketitle

\vspace{-1.5cm}

\begin{center}
  \rule{6cm}{1pt}
\end{center}

\abstract{We assess the possibilities offered by Hilbert space fundamentalism, an attitude towards quantum physics according to which all physical structures (\textit{e.g.} subsystems, locality, spacetime, preferred observables) should emerge from minimal quantum ingredients (typically a Hilbert space, Hamiltonian, and state). As a case study, we first mainly focus on the specific question of whether the Hamiltonian can uniquely determine a tensor product structure, a crucial challenge in the growing field of quantum mereology. The present paper reviews, clarifies, and critically examines two apparently conflicting theorems by Cotler \textit{et al.} and Stoica. We resolve the tension, show how the former has been widely misinterpreted and why the latter is correct only in some weaker version. We then propose a correct mathematical way to address the general problem of preferred structures in quantum theory, relative to the characterization of emergent objects by unitary-invariant properties. Finally, we apply this formalism in the particular case we started with, and show that a Hamiltonian and a state are enough structure to uniquely select a preferred tensor product structure.}
 

\vspace{0.4cm}
\hrule

\setcounter{tocdepth}{2}
\tableofcontents

\section{Introduction}

A Hilbert space, on its own, is quite featureless. Any transformation preserving the vector-space structure and its inner product, \textit{i.e.} any unitary transformation, is a symmetry. This makes any unit vector as good as any other, and any orthonormal basis can be rotated into any other. Normally, we break this symmetry by specifying a set of operators or, equivalently, some orthonormal bases that have operational meaning. The position and momentum operators for a particle, the creation and annihilation operators for a field, the computational basis for a qubit, the choice of a decomposition of the Hilbert space into tensor factors associated with subsystems,  are such examples.

One may be more ambitious and ask: \textit{can we recover some of this structure with less?} This is the central question raised by Hilbert space fundamentalism (HSF), a research programme that aims to recover the usual structures used in quantum mechanics, including the physical space, by only specifying a Hilbert space $\H$, a Hamiltonian $\hat H$, and, optionally, a state $\ket\psi$. HSF is the starting point of the radical ‘many-worlds’ interpretation known as Mad-Dog Everettianism~\cite{carroll2019mad,carroll2022reality}.

The idea that interesting structures can emerge from \textit{some} basic quantum data has a long history and is appealing to other research programs, such as, the Hilbert space first approach~\cite{piazza2010glimmers}. It is central to mereology~\cite{tegmark2015consciousness, carroll2021quantum, zanardi2024operational, loizeau2025quantum, vanrietvelde2025partitions, soulas2025disentangling}, and is also motivated by the ‘looming big problem’ of decoherence~\cite{zurek1998decoherence}, which questions the apparent arbitrariness of the system-environment split $\mathcal{H} = \mathcal{H}_S \otimes \mathcal{H}_E$, hoping that the correct decomposition may already be latent in the data $(\mathcal{H}, \hat{H}, \ket{\psi}) $~\cite{carroll2021quantum,adil2024search}. Other motivations to pursue this research program are: the quest for a quantum theory in which spacetime is not a background structure, and more generally quantum gravity considerations, where the notions of locality or subsystems might only emerge at scales much larger than the Planck scale~\cite{piazza2010glimmers, marolf2015comments, de2022frontiers, franzmann2024or}; quantum reference frames, which hint at a frame-dependent notion of subsystems \cite{ali2022quantum, castro2025relative,hoehn2023quantum} (although it remains an open question whether this can be made background independent \cite{franzmann2024or}); or the measurement problem \cite{franzmann2024or,stoica2024observation,stoica2025makes}.

In~\cite{cotler2019locality}, Cotler \textit{et al.} have proved a notorious result of HSF: a Hamiltonian can be enough to impose a tensor product structure (TPS) on the Hilbert space. More precisely, the authors showed that requiring the Hamiltonian to be \emph{$K$-local} with respect to some TPS, \textit{i.e.} involving interactions among only $K$ subsystems at a time, is generically sufficient: for most Hamiltonians, $K$-locality allows to uniquely characterize an equivalence class of TPSs. This is exciting as a TPS induces the quantum mechanical notions of entanglement, subsystems, and locality~\cite{kennedy1995empirical,barrett2007information,dibiagio2025circuitlocalityrelativisticlocality,dibiagio2025simple}. Ever since, this result has been cited \cite{cao2018bulk, carroll2019mad, giddings2019quantum, carroll2022reality} as a hallmark of the HSF program. 

However, \emph{it has been mostly understood in an incorrect way}, as if $K$-locality could select a \textit{single} TPS instead of an equivalence class. In response, Stoica has revealed in a series of papers \cite{stoica20213d,stoica2021refutation,stoica2023no, stoica2023prince, stoica2024does} how the widespread interpretation of Cotler \textit{et al.}'s result cannot hold. This has led to a theorem apparently constraining the ambitions of HSF and, more generally, the emergence of preferred structures in quantum theory. Stoica showed in particular that any $K$-local Hamiltonian admits an abundance of distinct TPSs which make it $K$-local, and that replacing the $K$-locality requirement with any other unitary-invariant condition would not change the situation.

In section~\ref{Hamil_TPS}, we first carefully review Cotler \textit{et al.}'s result and make explicit its definitions and assumptions; we then revisit and extend Stoica's arguments against HSF. Then, in the main section~\ref{characterizing}, we generalise from the Cotler \textit{et al.}-Stoica debate in order to extend the scope of the discussion. We first show that the apparent incompatibility between their conclusions dissolves once one distinguishes between absolute and relational notions of uniqueness, and argue that the latter is the suitable one in the context of quantum foundations (section~\ref{uniqueness}). Taking inspiration from invariant theory, we formalize this notion in high generality, making it applicable in principle to any kind of structure, providing insights on how structures mutually acquire their identity in a geometrical space (section~\ref{insights}). Equipped with this formalism, we can finally turn back to the initial problem and prove that a pair $(\hat{H}, \ket{\psi})$ can uniquely determine a TPS, contradicting one of Stoica's strongest claims~\cite{stoica20213d} (section~\ref{sec:maintheorem}). This refines the ambitions of HSF as well as clarifies the precise sense in which a notion of locality and subsystem structure can be said to \emph{emerge} from within a Hilbert space.

\section{Can the Hamiltonian uniquely determine a TPS?} \label{Hamil_TPS}

\subsection{Cotler \textit{et al.}'s theorem} \label{cotler}

Here are the definitions required to formulate Cotler \textit{et al.}'s result. Throughout this paper, an integer $n \geq 2$, some dimensions $(d_i)_{1 \leq i \leq n}$ with $d_i \geq 2$ and a Hilbert space $\H$ of dimension $\prod_{i=1}^n d_i$ are fixed once and for all. All the results in the sequel are restricted to the finite-dimensional case.

\begin{definition}[Tensor product structure] \label{def_TPS}
A \textit{TPS} of $\mathcal{H}$ is an equivalence class of isomorphisms $T : \mathcal{H} \rightarrow \bigotimes_{i=1}^n \mathcal{H}_i$ that factorize $\mathcal{H}$ into $n$ factors $\mathcal{H}_i$ of respective dimensions $d_i$, where two isomorphisms $T_1$ and $T_2$ are said to be equivalent (denoted $T_1 \sim T_2$) if $T_1 T_2^{-1}$ is a product of local unitaries $U_1 \otimes \dots \otimes U_n$ and arbitrary permutations of the factors.
\end{definition}

Once a TPS is fixed, the equivalence class $\mathcal T=[T]$ of isomorphisms allows to identify $\H$ with $\bigotimes_{i=1}^n \mathcal{H}_i$. It then makes sense to talk about single-site operators. Given, for each $i$, a choice $(O_i^\alpha)_{0 \leq \alpha \leq d_i^2-1}$ of an orthonormal\footnote{With respect to the Hilbert-Schmidt scalar product $\langle O , O'\rangle_\mathrm{HS} = \tr(O^\dagger O').$} basis for $\mathcal{L}(\mathcal{H}_i)$ with $O_i^0 = \bbone$, any Hermitian operator $\hat{H}$ can be uniquely decomposed as: 

\begin{equation}
\hat{H} = a_0 \bbone + \sum_{i=1}^n \sum_{\alpha \neq 0} a_i^\alpha \; O_i^\alpha + \sum_{1 \leq i < j \leq n} \; \sum_{\alpha,\beta \neq 0} a_{i j}^{\alpha \beta} \; O_i^\alpha O_j^\beta + \sum_{1 \leq i < j  < k \leq n} \; \sum_{\alpha, \beta, \gamma \neq 0} a_{i j k}^{\alpha \beta \gamma} \; O_i^\alpha O_j^\beta O_k^\gamma + \dots , \label{local_decomposition}
\end{equation}
where an operator like $O_i^\alpha$ is implicitly understood as acting as the identity on all factors except on $\mathcal{H}_i$. 
Note that there are $\prod_{i=1}^n d_i^2 = \dim(\mathcal{H})^2$ complex coefficients $a_{i j k\dots}^{\alpha \beta \gamma \dots}$, for \eqref{local_decomposition} is nothing else than the decomposition of $\hat{H}$ in the orthonormal basis\footnote{Since $\hat H$ is Hermitian, instead of choosing some bases $(O_i^\alpha)_\alpha$ of $\mathcal{L}(\mathcal{H}_i)$, it would have been enough to pick some bases of the Hermitian operators in $\mathcal{H}_i$, seen as a real vector space of dimension $d_i^2$, as are the Pauli or Gell-Mann matrices for example. Then, the expansion would require $\prod_{i=1}^n d_i^2$ \textit{real} coefficients $a_{i j k\dots}^{\alpha \beta \gamma \dots}$.} $(O_1^{\alpha_1}\cdots O_n^{\alpha_n})_{0 \leq \alpha_i \leq d_i^2-1}$ of $\mathcal{L}(\mathcal{H})$.

Hamiltonians encountered in physics are generally expressed in a relatively simple form, involving only interaction terms between a small number of subsystems. This idea is captured by the following definition.

\begin{definition}[$K$-locality] \label{def_locality}
Let $\mathcal{T}$ be a TPS of $\mathcal{H}$ into $\bigotimes_{i=1}^n \mathcal{H}_i$, and $\hat{H}$ be a Hermitian operator on $\mathcal{H}$ called Hamiltonian. For $K \in \{1,\dots,n\}$, we say that $\hat{H}$ is \textit{$K$-local} with respect to the TPS $\mathcal{T}$, or that the pair $(\hat{H}, \mathcal{T})$ is $K$-local, if there exists a choice of orthonormal bases $(O_i^\alpha)_{i,\alpha}$ for which the above decomposition \eqref{local_decomposition} of $\hat{H}$ involves only products of at most $K$ nontrivial operators (\textit{i.e.}~only the first $K$ sums have nonzero coefficients). Finally, $\hat{H}$ is \textit{$K$-local} if there exists a TPS $\mathcal{T}$ such that $(\hat{H}, \mathcal{T})$ is $K$-local. 
\end{definition}

Note that this definition is consistent because it does not depend on the particular choice of representative in the equivalence class $\mathcal{T}$. Acting with a unitary $U_i$ on each factor and permuting the indices $i \mapsto \sigma(i)$ still allows to choose the bases $\smash{(U_{\sigma(i)} O_{\sigma(i)}^k U_{\sigma(i)}^\dagger)}$ for which the coefficients $\smash{a_0, a_i^k, a_{i j}^{\alpha \beta}\dots}$ will be the same as previously, showing that $\hat{H}$ is indeed $K$-local for any other element in the equivalence class of the TPS.

\begin{remark} \textbf{\textit{K}-locality \textit{vs} spacetime locality.}
Although the above definition seems very natural, one could still question whether the notion of $K$-locality captures anything at all of the usual spacetime notion of locality. Consider, for instance, a variant of the 1D nearest-neighbour Ising model, where we add to the normal Hamiltonian some interaction terms of the form $\sigma^z_i \sigma^z_{i+1000}$: the resulting Hamiltonian is still 2-local, but would one say that it describes any local physics? In the same vein, the Newtonian or Coulombian interactions are clearly 2-local and yet they are known as infinite-range interactions. Conversely, a quantum field theory involving interactions with arbitrary many fields will still be spacetime local as long as the fields interact pointwise. Besides, it could be argued that \emph{strict} $K$-locality is not a very realistic physical requirement, and should rather be replaced by a notion of approximate $K$-locality. It has recently been shown, for instance, that most Hamiltonians  admit a TPS in which they are \textit{approximately} 2-local \cite{loizeau2023unveiling}.
\label{rem:locality}
\end{remark}

At this point, \emph{crucially} (but easily overlooked!), Cotler \textit{et al.} introduce the following equivalence relation, implementing the idea that a TPS acquires its identity only in relation to the other relevant structures involved, in this case the Hamiltonian (more on this in section~\ref{characterizing}).

\begin{definition}[Global unitary equivalence] \label{equivalence}
Two pairs $(\hat{H}, \mathcal{T})$ and $(\hat{H}', \mathcal{T}')$ are said equivalent if there exists a unitary $U \in \mathcal{U}(\mathcal{H})$ such that\footnote{To be precise, Cotler \textit{et al.} also allow for the transposed condition $\hat{H}' = U \hat{H}^{T} U^{\dagger}$ in their definition. See \cite[section 4.2]{cotler2019locality} for a discussion about the necessity of the transposition in the definition. \label{transpose}} $\hat{H}' = U \hat{H} U^{\dagger}$ and $T \sim T' U$, where $[T] = \mathcal{T}$ and $[T'] = \mathcal{T}'$. When $\hat{H}=\hat{H}'$, this definition reduces to\footnote{The fact that these two definitions are equivalent is claimed by Cotler \textit{et al.}, and will be proved in full generality in section \ref{characterizing}. \label{twofold_definition}} an equivalence relation between TPSs. In this case, we say that $\mathcal{T}$ and $\mathcal{T}'$ are \textit{equivalent with respect to $\hat{H}$} if $T \hat{H}T^{-1}$ and $T' \hat{H}T'^{-1}$ on $\bigotimes_{i=1}^n \mathcal{H}_i$ are the same up to single-site unitaries and permutations of factors.
\end{definition}

\noindent Intuitively, this equivalence relation allows us to consider classes of TPSs in which $\hat{H}$ looks the same (at least from the point of view of unitary-invariant properties). From there, Cotler \textit{et al.} define the notion of $K$-dual TPSs.

\begin{definition}[$K$-duality] \label{k-duality}
Given a Hamiltonian $\hat{H}$, we say that two TPSs are \textit{$K$-duals} if they are not equivalent with respect to $\hat{H}$ (according to definition \ref{equivalence}) but $\hat{H}$ is $K$-local in both TPSs.
\end{definition}

\noindent An example of dual TPSs is given by the Jordan-Wigner transformation on the 1D Ising model, which provides two sets of variables for which the Hamiltonian takes a 2-local form, and yet these variables are related in a nonlocal way so that the two TPSs are inequivalent\footnote{If $\hat{H}=J\sum_i\sigma_z^{(i)}\sigma_z^{(i+1)}+h\sum_i\sigma_x^{(i)}$ denotes the 2-local Hamiltonian of a chain of spin-1/2 systems, it turns out that the change of variables $\mu_z^{(i)}=\prod_{j\leq i}\sigma_x^{(j)}$; $\mu^{(i)}_x=\sigma^{(i)}_z\sigma^{(i+1)}_z$; $\mu^{(n)}_x=\mu^{(n)}_z$ yields an inequivalent TPS in which the Hamiltonian still takes a 2-local form $\hat{H}= J\sum_i\mu_x^{(i)} + h\sum_i \mu_z^{(i)}  \mu_z^{(i+1)} - J\mu_x^{(n)} + h\mu_z^{(1)}$~\cite[section 1]{cotler2019locality}.}.

For a Hamiltonian, being $K$-local with respect to a TPS with $n$ factors for some $K \ll n$ is a very particular property that will in general not be satisfied. However, when it holds, is the existence of duals the rule or the exception (possibly due to strong symmetries in the case of the Ising model)? The question addressed by Cotler \textit{et al.} is the following: given a Hamiltonian and a TPS that makes it $K$-local for some small $K$, does this TPS have any $K$-duals? Here is the main theorem proved in \cite{cotler2019locality}:

\begin{theorem}[Cotler \textit{et al.}]
Assume $d_1 = \dots = d_n \equiv d$, and suppose the existence of a single Hamiltonian on $\mathcal{H}$ admitting a TPS $\H \rightarrow \bigotimes_{i=1}^n \mathcal{H}_i$ that makes it $K$-local but without any $K$-duals. Then, if $K$ is sufficiently small, almost all $K$-local Hamiltonians on $\H$ do not have any $K$-dual TPS.
\label{cotler_claim}
\end{theorem}

\noindent Here, ‘almost all’ means ‘for all except a measure-zero set’, where the measure-zero sets are defined with respect to the Lebesgue measure on Hermitian matrices, which coincide with several other measures such as the well-studied Gaussian unitary ensemble \cite[footnote 8]{cotler2019locality}.

The proof is intricate and involves highly nontrivial results from algebraic geometry. Moreover, the necessity to rely on the assumption that there exists at least one Hamiltonian without any duals is an important weakness of the result, acknowledged by the authors. To justify the validity of the assumption, they provide numerical simulations that seem to indicate the existence of such Hamiltonians with no $K$-dual TPSs in some tractable cases. If they do exist in general in interesting cases, it follows indeed that $K$-locality allows to select a unique \emph{equivalence class} of TPSs (in the sense of definition \ref{equivalence}).

\subsection{Stoica's theorem} \label{stoica}

Let's now turn to Stoica's seemingly incompatible result. In \cite{stoica20213d, stoica2024does}, the following theorem is proved.

\begin{theorem}[Stoica] \label{theorem_stoica}
If a Hamiltonian $\hat H$ admits a TPS in which it is $K$-local, then such a TPS is not unique.
\end{theorem}

\noindent Importantly, Stoica does not consider here the structures in Hilbert space up to the global unitary equivalence of definition \ref{equivalence}, so that ‘unique’ must be understood in a strict sense. In the rest of the paper, we will be careful to distinguish between the words ‘distinct’ (Stoica's notion of uniqueness) and ‘inequivalent’ (Cotler \textit{et al.}'s notion). We will later examine, in section \ref{uniqueness}, which is the appropriate notion of uniqueness in the context of quantum foundations.

Stoica's theorem follows immediately from the following two lemmas. In the sequel, $U \cdot \mathcal{T}$ denotes the equivalence class $[TU]$ where $[T] = \mathcal{T}$, according to the equivalence relation of definition \ref{def_TPS}. Note that this object is well-defined, meaning that it is independent of the particular choice of representative $T$, because $T_1 \sim T_2 \Rightarrow T_1U \sim T_2U$ as $T_1T_2^{-1}=(T_1U)(T_2U)^{-1}$.

\begin{lemma}\label{K-locality-lemma}
Let $U$ be a unitary operator on $\mathcal{H}$. The pair $(\hat{H}, \mathcal{T})$ is $K$-local if and only if the pair ${(U \hat{H} U^\dagger ,  U \cdot \mathcal{T})}$ is $K$-local.
\end{lemma}

\begin{lemma}\label{symmetries-lemma}
For any $K$-local pair $(\hat{H}, \mathcal{T})$, there exists at least one unitary $U$ such that $U \hat{H} U^\dagger=\hat H$ but $U \cdot \mathcal{T} \neq \mathcal T$.
\end{lemma}

\noindent Given these results, starting from a TPS in which $\hat{H}$ is $K$-local, one can always build a distinct TPS that has the same property. Indeed, let $(\hat{H}, \mathcal{T})$ be a $K$-local pair. Then by lemma~\ref{symmetries-lemma}, there exists a unitary $U$ such that  $U \hat{H} U^\dagger = \hat H$ but $U \cdot \mathcal{T} \neq \mathcal T$. By lemma~\ref{K-locality-lemma}, then, it follows that $(\hat H, U \cdot \mathcal{T})$ is another distinct $K$-local pair.

Note, however, that $\mathcal{T}$ and $\mathcal{T}'$ will not be $K$-duals in the terms of definition \ref{k-duality}, because the two pairs are unitarily equivalent. Here we can already start to see the reconciliation between the Cotler \textit{et al.} and Stoica results, as the non-uniqueness implied by theorem \ref{theorem_stoica} is precisely what defines the equivalence class singled out by $K$-locality. Impatient readers can refer to section~\ref{characterizing} where we deepen this discussion, while the rest of this section revisits the proof of Stoica's theorem.

\subsubsection{$K$-locality is unitary-invariant} \label{invariant}

The proof of lemma~\ref{K-locality-lemma} is immediate and follows from the fact that the coefficients $a_{i j k\dots}^{\alpha \beta \gamma \dots}$ in \eqref{local_decomposition} are nothing else than the Hilbert-Schmidt scalar products:
\begin{equation}
a_{i j k\dots}^{\alpha \beta \gamma \dots} = \braket{\hat{H} , O_i^\alpha O_j^\beta O_k^\gamma\cdots }_\text{HS}.
\end{equation}
Since the Hilbert-Schmidt scalar product is invariant under unitary transformations, we have
\begin{equation}
 \braket{U \hat{H} U^\dagger , U [O_i^\alpha  O_j^\beta  O_k^\gamma  \cdots ] U^\dagger}_\text{HS}= \braket{\hat{H} , O_i^\alpha O_j^\beta  O_k^\gamma\cdots }_\text{HS}=a_{i j k\dots}^{\alpha \beta \gamma \dots}.
\end{equation}
Therefore, the decomposition of $\hat{H}$ relative to the basis of operators $(O_i^\alpha O_j^\beta O_k^\gamma \dots)_{i,\alpha, j, \beta, k, \gamma \dots}$ involves exactly the same coefficients as the decomposition of $U \hat{H} U^\dagger$ in $U \cdot \mathcal{T}$ relative to the basis $(U [O_i^\alpha O_j^\beta O_k^\gamma \dots] U^\dagger)_{i,\alpha, j, \beta, k, \gamma \dots}$. If only the first $K$ sums in \eqref{local_decomposition} contain nonvanishing coefficients, this is also the case for $U \hat{H} U^\dagger$ decomposed in the transformed basis.

\medskip

There are several ways of proving lemma~\ref{symmetries-lemma}, and this is what we will focus on next.

\subsubsection{$\hat{H}$ has more symmetries than $\mathcal{T}$} \label{symmetries}

Lemma~\ref{symmetries-lemma} is derived in \cite{stoica2024does} by a simple dimension-counting argument. We reproduce and clarify the proof below.

\begin{proposition} \label{more_symmetries}
Let $\hat{H}$ be a Hamiltonian and $\mathcal{T}$ a TPS of $\mathcal{H}$. Then there are symmetries of $\hat H$ that are not symmetries of $\mathcal T$. That is, the stabilizer subgroups $\stab(\hat{H})={\{ U \in \mathcal{U}(\mathcal{H}) \mid U \hat{H} U^\dagger = \hat{H} \}}$ and $\stab(\mathcal T)={\{ U \in \mathcal{U}(\mathcal{H}) \mid U \cdot \mathcal{T} = \mathcal{T} \}}$  are such that
\[ \stab(\hat{H}) \not\subset \stab(\mathcal{T}).\]
\end{proposition}

\begin{proof}
Let $\mathcal{B}_{\hat{H}}$ be an eigenbasis of $\hat{H}$. The set $\Gamma$ of unitaries that are diagonal in $\mathcal{B}_{\hat{H}}$ is a commutative Lie subgroup of $\stab(\hat{H})$ of real dimension $\dim \mathcal{H}$, because $\Gamma$ is exactly the set of operators diagonal in $\mathcal{B}_{\hat{H}}$ with eigenvalues of the form $e^{i\theta}$ for some $\theta \in \mathbb{R}$.

On the other hand, $\stab(\mathcal{T})$ is by definition the set of local unitaries $U_1 \otimes \dots \otimes U_n$ plus possible permutations. Any commutative Lie subgroup of $\stab(\mathcal{T})$ is composed of tensor products of co-diagonalizable unitaries in each factor (with possibly even more restrictions if some permutations are allowed). There are at most $d_i$ real parameters for the eigenvalues in each factor, but one has to take care of the fact that, in a TPS, any phase multiplication within one factor factorizes out as a global multiplication. Hence, any commutative Lie subgroup of $\stab(\mathcal{T})$ has maximal real dimension
\begin{equation}
    D \leq \underbrace{\dim_\mathbb{R} \mathrm{U}(1)}_\text{global} + \sum_{i=1}^n \Big( d_i - \underbrace{\dim_\mathbb{R} \mathrm{U}(1)}_\text{local} \Big) = \sum_{i=1}^n d_i - (n-1) \,.
\end{equation}
Furthermore, recalling that $n \geq 2$ and $d_i \geq 2$ for all $i$, the inequality $d_1 \cdots d_n > \sum_{i=1}^n d_i -( n-1 )$ holds. This can be shown by induction. For $n=2$, assume without loss of generality that $d_2 = \min(d_1 , d_2)$, and observe that $d_1 d_2 \geq 2 d_1 > d_1 + d_1 - 1 \geq d_1 + d_2 -1$. Now, if the property holds for $n-1$ integers, assume without loss of generality that $d_n = \min(d_1, \dots , d_n)$, and write:
\begin{equation}
\begin{aligned}
    d_1 \cdots d_n 
    &\geq 2 d_1 \cdots d_{n-1} \\
    &> 2\big(d_1 + \dots + d_{n-1} - (n-2)\big) \\
    &>  d_1 + \dots + d_{n-1} + \underbrace{d_1 + \dots + d_{n-1}}_{\geq  d_n + 2(n-2)} - 2n + 4 \\
    &>\smash{\sum_{i=1}^n} d_i,
\end{aligned}
\end{equation}
which entails \textit{a fortiori} the desired inequality. Consequently, if we had $\stab(\hat{H}) \subset \stab(\mathcal{T})$, then $\Gamma$ would be a commutative Lie subgroup of $\stab(\mathcal{T})$ of dimension $\dim \H=d_1 \cdots d_n$, which is impossible. 
\end{proof}

\noindent Remark that the mere observation of $\dim_\mathbb{R} \stab(\mathcal{T}) = \sum_{i=1}^n d_i^2 - n + 1$ and ${\dim_\mathbb{R} \stab(\hat{H}) \geq d_1 \cdots d_n}$ would already have led to a contradiction for $n$ large enough, because the latter grows asymptotically much faster than the former. This implies that, for $n$ large enough, almost all symmetries of $\hat{H}$ are not symmetries of $\mathcal{T}$. Indeed, $\stab(\hat{H}) \cap \stab(\mathcal{T})$ is in this case a submanifold of $\stab(\hat{H})$ of strictly smaller dimension, so only a measure zero subset of $\stab(\hat{H})$ are also symmetries of $\mathcal{T}$. Stoica's argument would work for any kind of structure $\mathcal{S}$ (not necessarily a TPS) whose group of symmetries forms a Lie group and whose dimension can be strictly bounded by that of $\stab(\hat{H})$.

This completes the proof of theorem \ref{theorem_stoica}, but Stoica provides another illuminating reason why lemma~\ref{symmetries-lemma} holds, which we examine next.

\subsubsection{The time evolution symmetry} \label{time}

 Given $\hat{H}$, we readily know a one-parameter family of symmetries of $\hat{H}$: the time evolution unitaries $(e^{-it\hat{H}})_{t \in \mathbb{R}}$. This is interesting because, from a physical point of view, we know that in realistic situations these operators will not fix the TPS; otherwise, this would yield a theory in which no entanglement ever occurs. Mathematically, we can easily check that restricting our study to those operators is sufficient to prove the lemma, at least when $\hat{H}$ is not 1-local, due to the following equivalence.

\begin{proposition} \label{lemma_time}
The time evolution unitaries $(e^{-it\hat{H}})_{t \in \mathbb{R}}$ are all symmetries of $\mathcal{T}$ if and only if $\hat{H}$ is 1-local with respect to $\mathcal{T}$.
\end{proposition}

\begin{proof}
Suppose $e^{-it\hat{H}} \in \stab(\mathcal{T})$ for all $t$, meaning that it can be written in the form $(U_1(t) \otimes \dots \otimes U_n(t)) \circ U_\sigma(t)$. By continuity of the map $t \mapsto (U_1(t) \otimes \dots \otimes U_n(t)) \circ U_\sigma(t)$, and considering that only the $U_\sigma$ part can possibly permute factors, along with $U_\sigma(0) = \bbone$, we deduce that this permutation is continuously connected to the identity, hence $U_\sigma(t) = \bbone$ for all $t$. Now, differentiating the equality $e^{-it\hat{H}} = U_1(t) \otimes \dots \otimes U_n(t)$ with respect to $t$ at $t=0$ yields $\hat{H} = A_1 \otimes \bbone \otimes … \otimes  \bbone + … + \bbone \otimes … \otimes \bbone \otimes A_n$ for some Hermitian operators $A_i$, therefore $\hat{H}$ is 1-local. Conversely, if $\hat{H}$ is 1-local for $\mathcal{T}$, $e^{-it\hat{H}}$ is clearly a product of local unitaries.
\end{proof}

Consequently, since 1-local Hamiltonians form a strictly lower dimensional subspace of the space of $K$-local Hamiltonians, most Hamiltonians will generate at least one unitary $U = e^{-it\hat{H}}$ for some $t \in \mathbb{R}$ that is not a symmetry of $\mathcal{T}$, while it is obviously a symmetry of $\hat{H}$. Following the method outlined at the beginning of section \ref{stoica}, this allows us to construct, for most $K$-local Hamiltonians, distinct TPSs in which $\hat{H}$ is $K$-local.

\begin{remark} \textbf{Taking the Heisenberg picture seriously.} \label{remark:descriptors}
The overall idea behind this argument appears clearly in the light of Zanardi \textit{et al.}'s theorem \cite{zanardi2004quantum}. While a TPS $\H \rightarrow \bigotimes_{i=1}^n \H_i$ obviously induces a decomposition of the algebra $\L(\H)$ as a product of the algebras over the different factors, namely $\L(\H)\cong \bigotimes_{i=1}^n \L(\H_i)$, the converse is also true for finite-dimensional Hilbert spaces. Specifically, let $(\A_i)_{1\leq i \leq n}$ be a family of subalgebras of $\L(\H)$ such that: 
\begin{enumerate}
\item the $\A_i$'s pairwise commute;
\item they have trivial intersection $\A_i \cap \A_j = \{ \bbone \}$ for $i\neq j$; 
\item they generate the whole $\L(\H)$;
\end{enumerate}
then there exists a unique TPS $\H \to \bigotimes_{i=1}^n \H_i$ such that $\L(\H_i) \cong \A_i$ for all $i$. Physically, this translates the fact that one can specify the subsystems of a physical system by specifying their algebras of observables, and vice-versa. This algebraic viewpoint underlies the \emph{descriptor formulation} of quantum theory~\cite{deutsch2000information,bedard2021abc}, also known as the Deutsch-Hayden formulation. In a slogan, the descriptors theory is the Heisenberg picture taken seriously: its main idea is that a \emph{complete and local} description of any quantum system is given by its evolving algebra of observables. The Heisenberg picture time evolution twists these subalgebras and therefore the TPS into a new one (corresponding to entanglement generation in the Schrödinger picture), although the algebraic relations between observables are preserved at all times. In particular, for all $t$, $[\mathcal{A}_i(t),\mathcal{A}_j(t)]=0~~~\text{for }i\neq j,$ and $\bigvee_{i=1}^n \mathcal{A}_i(t)=\mathcal{L}(\mathcal{H})$ (see \cite{andreadakis2025tensor} for a geometrical description of how a TPS gets twisted in time). 

The Schrödinger and the Heisenberg pictures are usually deemed equivalent, but it is not entirely correct. Intuitively, this should not be so surprising, as the dimension of the set of descriptors in $\H$ is larger than the dimension of the set of density matrices in $\H$. In particular, the two pictures become non-equivalent when $\H$ is equipped with a TPS $\H \to \bigotimes_{i=1}^n \H_i$. Contrary to the map $\ket \psi \mapsto (\rho_i)_{1 \leq i \leq n}$ associating to any state in $\H$ the list of partial states in the $\H_i$'s, the map $\ket \psi \mapsto (q_i)_{1 \leq i \leq n}$ associating the list of partial descriptors is injective. This means that, in the Heisenberg picture, no information is lost when going from the full state to the partial states, and that the former can be reconstructed from the latter. Here lies the fundamental difference between descriptors and density matrices, which makes descriptor theory so appealing.
\end{remark}

\subsubsection{A continuous infinity of distinct TPSs} \label{infinity}

The time evolution argument above does not only imply the existence of many distinct TPSs that make the Hamiltonian $K$-local, but it also leads to the existence of a \textit{continuous} orbit of such TPSs. This result is not particularly important in itself for our present purposes, but the method of the proof is insightful and introduces some of the ideas of section \ref{characterizing}.

\begin{proposition} \label{infinity}
Suppose $\hat{H}$ is not 1-local with respect to some TPS $\mathcal{T}_0$. Then the time evolution unitaries $(e^{-it\hat{H}})_{t \in \mathbb{R}}$ acting on $\mathcal{T}_0$ generate an uncountable infinity of \emph{different} TPSs.
\end{proposition}

\begin{proof}
Denote $U_t = e^{-it\hat{H}}$ and $(\mathcal{T}_t)_{t \in \mathbb{R}} = (U_t \cdot \mathcal{T}_0)_{t \in \mathbb{R}}$ the orbit of $\mathcal{T}_0$ under the action of the time-evolution operators. The strategy of the proof is to build a function on the set of all TPSs of $\mathcal{H}$ (in particular, it should be compatible with its quotient structure, \textit{i.e.} unaffected by local unitaries), that maps any TPS $\mathcal{T}$ to a real number $\varphi(\mathcal{T}) \in \mathbb{R}$, such that $t \mapsto \varphi(\mathcal{T}_t)$ is continuous but not constant over time. It will then follow that the latter takes an uncountable infinity of different values; consequently, there must be an uncountable infinity of distinct TPSs in $(\mathcal{T}_t)_{t \in \mathbb{R}}$.

Here is a possible way to build $\varphi$. Since $\hat{H}$ is not 1-local with respect to $\mathcal{T}_0$, there must exist a state $\ket{\psi} \in \mathcal{H}$ such that $\ket{\psi}$ is a product state in $\mathcal{T}_0$, but gets entangled in $\mathcal{T}_t$ under the time evolution (this is simply the Heisenberg picture). Precisely, this means the following. Pick $T_0 \in \mathcal{T}_0$ a representative of $\mathcal{T}_0$, and consider $T_t \equiv T_0 \circ U_t  \in \mathcal{T}_t$; these are two isomorphisms from $\mathcal{H}$ to $\bigotimes_{i=1}^n \mathcal{H}_i$. Denote by $\tr_{\bar{i}}$ the partial trace over all factors of $\otimes_{i=1}^n\H_i$ except the $i^\text{th}$ and $S(\rho)$ the von Neumann entropy of a density matrix $\rho$. Now, one can find a $\ket{\psi}$ and an index $i$ such that $S\left( \tr_{\bar{i}}( T_t \ket{\psi}\!\!\bra{\psi} T_t^ {-1}) \right)$ equals $0$ at $t=0$ but is positive for some $t>0$. Indeed, if such a $\ket{\psi}$ didn't exist, this would imply that all product states remain product states in $\mathcal{T}_0$ under the time evolution, which contradicts the fact that $\hat{H}$ is not 1-local in this TPS (recall proposition \ref{lemma_time}).

The function $\varphi$ can then be defined as $\varphi : [T] \mapsto S\left( \tr_{\bar{i}}( T \ket{\psi}\!\!\bra{\psi} T^ {-1}) \right)$. It is well-defined, namely it does not depend on the choice of representative $T$, because the possible permutation can again be ignored due to the continuity of $t \mapsto U_t$, and because for all $A \in \mathcal{L}(\mathcal{H})$ we have
\begin{equation}
\tr_{\bar{i}}\left( U_1 \otimes \dots \otimes U_n \; A \; U_1^\dagger \otimes \dots \otimes U_n^\dagger  \right) = U_i \tr_{\bar{i}}(A) U_i^\dagger,
\end{equation}
along with the fact that $S$ is a unitary-invariant function satisfying $S(U \rho U^\dagger) = S(\rho)$. Finally, $t \mapsto \varphi(\mathcal{T}_t)$ is clearly continuous.
\end{proof}

Of course, the von Neumann entropy is simply a convenient choice here. In fact, any continuous unitary-invariant real function quantifying entanglement, like the purity $\tr(\rho^2)$, would equally work.

\section{How structures see each other} 
\label{characterizing}

\subsection{Which notion of uniqueness to keep?} \label{uniqueness}

At first glance, the two theorems presented in sections~\ref{cotler} and \ref{stoica} seem incompatible. Does the Hamiltonian allow to uniquely select a TPS \textit{via} the $K$-locality requirement (as suggested by Cotler \textit{et al.}'s theorem), or is it impossible (as suggested by Stoica's theorem)? In fact, the two claims are not contradictory: Cotler \textit{et al.}'s theorem only selects a TPS \emph{up to unitary equivalence} on the pair $(\hat{H}, \mathcal{T})$, while Stoica's construction outputs distinct TPSs (in the sense of definition \ref{def_TPS}) that are all clearly unitarily equivalent (in the sense of definition \ref{equivalence}) because they are generated by applying a global unitary. 

That said, what is the lesson to draw for the HSF program? Should we maintain hope to see our physical theories emerging from minimal quantum ingredients? To answer this, it is first important to observe that the two theorems illustrate two different attitudes towards mathematical structures. For Cotler \textit{et al.}, a structure acquires its identity only relative to other structures. This is the purpose of the equivalence relation of definition \ref{equivalence}: the quotient remembers only the unitary-invariant relations between $\hat{H}$ and $\mathcal{T}$ (it can distinguish \textit{e.g.} how entangled the eigenvectors of $\hat{H}$ are in $\mathcal{T}$) but washes out everything else (it can not discriminate between, say, $\mathcal{T}$ and $e^{-it\hat{H}} \cdot \mathcal{T}$ with respect to $\hat{H}$). For Stoica, on the other hand, the structures are perceived in a more absolute sense: $\mathcal{T}$ and $e^{-it\hat{H}} \cdot \mathcal{T}$ are seen as distinct TPSs, regardless of whether they can be distinguished by some surrounding structure. After all, a Hilbert space is a set (satisfying certain axioms), and a set is nothing but a list of objects that can be referred to and labeled. In this absolute perspective, there is a continuous infinity of different ways to choose a unit vector in a bare Hilbert space $\H$; whereas in the previous relational perspective, there is only one way to choose it.

The importance of this distinction appears clearly in view of the misuses of Cotler \textit{et al.}'s theorem, which has been interpreted as a tool to select a unique TPS \emph{in the absolute sense}; see, for instance, how the result is mentioned in \cite{cao2018bulk, carroll2019mad, giddings2019quantum, lucas2019operator, carroll2022reality}.
Carroll writes: ‘As shown by Cotler \textit{et al.}, generic Hamiltonians admit no local factorization at all, and when such a factorization exists, it is unique up to irrelevant internal transformations (and some technicalities that we won’t go into here). Therefore, the spectrum of the Hamiltonian is enough to pick out the correct notion of an emergent spatial structure when one exists’ \cite{carroll2022reality}. Yet, the internal transformations are irrelevant only if one focuses on the structure $(\hat{H}, \mathcal{T})$, but they can become relevant when one adds $\ket{\psi}$ or any other structure usually embedded in the theory. 
Consider for example the task of building an emergent spacetime from a fixed triple $(\H,\hat{H}, \ket{\psi})$  with the idea that the supposedly unique TPS provided by Cotler \textit{et al.}'s theorem is used in such a way that Hilbert space factors $\mathcal{H}_i$ will correspond to regions of space and then build a metric structure on top of that, based on the mutual information between the factors computed from $\ket{\psi}$. Different choices of TPSs \textit{within the equivalence class} provided by Cotler \textit{et al.}'s theorem will lead to different spacetime geometries, as $\ket{\psi}$ does not necessarily bear the same entanglement structure with the different representatives.

Is one of the two attitudes (relational \textit{vs.} absolute) more relevant in the quantum foundations context? We argue that \emph{the relevant notion of uniqueness in physics is relational}\footnote{The relational perspective is also reasonable from a mathematician's point of view, in particular since the advent of category theory, in which the identity of an object emerges from the arrows \textit{i.e.} from the relations that link it to itself and the others, as expressed by Yoneda's lemma.}, meaning that a preferred emergent structure should be unique up to a global unitary equivalence on the whole set of structures involved. Indeed, \emph{two sets of structures in a Hilbert space related by a global unitary will yield exactly the same theoretical predictions}, hence they depict one and the same physical theory\footnote{As remarked by Cotler \textit{et al.}, there is always another set of structures that generates the same predictions: the transposed theory (recall footnote \ref{transpose}). Rigorously speaking, the true group of symmetry that should be considered in this work may be $\mathcal{U}(\H) \times \mathbb{Z}/2\mathbb{Z}$, but this would not affect the results obtained here, because our discussion will be fully general and applicable in principle to any symmetry group.}; this idea famously motivates the Stone--von Neumann theorem. However, the equivalence relation of definition~\ref{equivalence} employed in~\cite{cotler2019locality} must be modified if, in addition to the Hamiltonian, a state $\ket{\psi}$ also plays a role, as illustrated in the previous paragraph.

\pagebreak

In conclusion, the correct mathematical way to address the question raised by HSF, concerning the uniqueness of emergent structures in quantum mechanics, is: 
\begin{enumerate}
\item specify the kind of the input structure $\mathcal{S}_0$ and the kind of the structure $\mathcal{S}_e$ that is desired to emerge from $\mathcal{S}_0$ in $\H$;
\item find a unitary-invariant property $P$ of the pair $(\mathcal{S}_0, \mathcal{S}_e)$ such that a pair satisfying $P$ is unique up to unitary equivalence, \textit{i.e.}: 
\begin{equation} \label{intuition}
    P(\mathcal{S}_0, \mathcal{S}_e) \text{ and } P(\mathcal{S}'_0, \mathcal{S}_e') \implies \exists\, U \in \mathcal{U}(\H): (\mathcal{S}'_0, \mathcal{S}_e') = U \cdot (\mathcal{S}_0, \mathcal{S}_e).
\end{equation}
\end{enumerate}
We shall now properly formalize this idea in the following section.

\subsection{Insights from invariant theory} \label{insights} 

To do so, let's now introduce the following crucial definition.

\begin{definition} \label{kind}
A \emph{kind} is a set $\mathcal{K}$ on which the unitary group $\mathcal{U}(\H)$ acts, and an element $\mathcal{S} \in \mathcal{K}$ is called a $\mathcal{K}$-structure. We say that $\mathcal{K}$ is a \emph{determined kind} if moreover this action is transitive, \textit{i.e.} if $\mathcal{K} = \mathcal{U}(\H) \cdot \{\mathcal{S}\}$ is composed of only one orbit.
\end{definition}

\noindent The notion of kind was introduced by Stoica~\cite{stoica20213d} as a general notion for talking about ‘what sort of object’ a structure is within the Hilbert space\footnote{Stoica's notion of kind, definition 3 in \cite{stoica20213d}, is more involved, requiring the $\mathcal K$-structures to satisfy certain unitary-invariant properties (a list of tensor equations and inequations). However, we find that for our purposes, all we need is the existence of an action of the unitary group.}. The concept is indeed quite flexible.

For instance, the set $\mathcal{K}_\text{vector}$ of all vectors in $\H$, the set $\mathcal{K}_\text{Herm}$ of all Hermitian operators, the set $\mathcal{K}_{\mathfrak{G}\text{-rep}}$ of all the representations on $\H$ of a Lie algebra $\mathfrak{G}$ or the set $\mathcal{K}_\text{TPS}$ of all tensor product structures for $\H$ are examples of kinds. Furthermore, the set $\mathcal{K}_{\text{vector}(1)}$ of all unit vectors in $\H$ or the set $\mathcal{K}_{\text{Herm}(\sigma)}$ of all Hermitian operators with given spectrum $\sigma$ (counting multiplicities) are determined kinds. Similarly, the set $\mathcal{K}_{\text{TPS}(n; d_1, \dots, d_n)}$ of all TPSs of $\H$ into $n$ factors of respective dimensions $(d_i)_{1 \leq i \leq n}$ is a determined kind, because a TPS $\mathcal{T}$ is fully characterized by a proper labeling of an orthonormal basis that is separable in $\mathcal{T}$, and the unitary group is transitive on the orthonormal bases. The Stone--von Neumann's theorem basically states that $\mathcal{K}_{\mathfrak{H}\text{-rep}}$, where $\mathfrak{H}$ denotes the Heisenberg Lie algebra, is a determined kind (up to some technicalities to deal with unbounded operators and domain issues). Given an integer $N$ and a $N \times N$ positive semidefinite matrix $G$, the set $\mathcal{K}_{N\text{-vectors}(G)} = \{ (\ket{\psi_i})_{1 \leq i \leq N} \mid \forall i,j, \quad  \braket{\psi_i \vert \psi_j} = G_{ij} \}$ of families of $N$ vectors with specified pairwise scalar products is also a determined kind\footnote{To see that, let $(\ket{e_i})_{1 \leq i \leq n}$ and $(\ket{f_j})_{1 \leq j \leq n}$ be two families of vectors. If they satisfy the property, their Gram matrices are identical. Applying the Gram-Schmidt procedure in $V = \mathrm{span}(\ket{e_i})_i$ and $W =  \mathrm{span}(\ket{f_j})_j$ yields two orthonormal bases $\mathcal{B}_V = (\ket{v_i})_i$ and $\mathcal{B}_W = (\ket{w_j})_j$ of $V$ and $W$, and for all $i$ the coefficients of $\ket{e_i}$ decomposed in $\mathcal{B}_V$ are the same as the coefficients of $\ket{f_i}$ decomposed in $\mathcal{B}_W$, because exactly the same procedure was applied in each subspace starting from the same Gram matrix. Let now $U$ be the unitary mapping $\mathcal{B}_V$ to $\mathcal{B}_W$. Since $U$ preserves the projections on the basis elements, we have $U \ket{e_i} = \ket{f_i}$ for all $i$.}. For the particular choice $G=\bbone$, this simply yields the kind of orthonormal bases in $\H$.

If $\mathcal{K}_0$ and $\mathcal{K}_e$ are two kinds, then their Cartesian product $\mathcal{K}_0 \times \mathcal{K}_e$ is still a kind. However, the product of two determined kinds is not necessarily determined. One can, however, try to put restrictions on the product kind in order to make it determined. Let's emphasise an important example below.
\begin{example} \label{kind_HSF}
For example, the kinds $\mathcal K_{\text{Herm}(\sigma)}$ and $\mathcal K_{\text{vector}(1)}$ are both determined kinds, while their product is not determined\footnote{Note that $\bra{\psi} \hat H\ket \psi =(\bra{\psi}U^\dagger) (U\hat HU^\dagger)) U\ket \psi$ for all unitaries $U$, while $\mathcal K_{\text{Herm}(\sigma)}\times \mathcal K_{\text{vector}(1)}$ contains pairs $(\hat H,\ket\psi)$ and $(\hat H,\ket{\phi})$ such that $\bra\psi\hat H\ket\psi\neq\bra\phi\hat H\ket\phi$ so the action of $\mathcal U(\H)$ is not transitive on this product kind.}. However, we can define a restricted kind $\mathcal{K}_{\text{HSF}(\sigma,\Lambda )}\subset\mathcal K_{\text{Herm}(\sigma)}\times \mathcal K_{\text{vector}(1)}$ that is determined, as follows. Let $(\Pi_i^{\hat{H}})_{1\leq i \leq n}$ denote the eigenprojectors of $\hat{H}$ and $\Lambda = (\lambda_i)_{1\leq i \leq n}$ a family of non-negative real numbers such that $\sum_i \lambda_i = 1$. Then the set of Hermitian operators with given spectrum and states with given modulus square projections on its eigenspaces:
\begin{equation}
\mathcal{K}_{\text{HSF}(\sigma,\Lambda )} = \{ (\hat{H} , \ket{\psi}) \mid \mathrm{Spec}(\hat{H}) = \sigma \text{ and } \forall i, \; \lVert \Pi_i^{\hat{H}} \ket{\psi} \rVert^2 = \lambda_i\}
\end{equation}
is a determined kind. To see this, observe that one can always find a unitary mapping $\hat{H}$ to $\hat{H}'$ if they have identical spectrum, and one can still apply any complex phase rotation along the eigenvectors, sufficient to attain any $\ket{\psi'}$ with the same modulus square projections in the eigenbasis. Besides, note that, in this case, the action of $\mathcal{U}(\H)$ is moreover free because $\hat{H}'$ determines the unitary applied up to phases in its eigenbasis, which are exactly encoded in $\ket{\psi'}$.
\end{example}

With this in mind, we can now give a general definition of the notion of a preferred structure uniquely characterized (in the relational sense) by some property, as motivated in the previous section.

\begin{definition}[Relational uniqueness] \label{P_determines}
Let $\mathcal{K}_0$ and $\mathcal{K}_e$ be two determined kinds, and $P$ a unitary-invariant property on the product kind $\mathcal{K}_0 \times \mathcal{K}_e$. We say that $P$ \emph{determines} the product kind if $P$ holds on exactly one orbit in $\mathcal{K}_0 \times \mathcal{K}_e$ under the action of $\mathcal{U}(\H)$. In other words, the set $\{ (\mathcal{S}_0 , \mathcal{S}_e)\in \mathcal{K}_0 \times \mathcal{K}_e \mid P(\mathcal{S}_0 , \mathcal{S}_e) \}$ must be non-empty and:

\begin{equation}
P(\mathcal{S}_0 , \mathcal{S}_e) \text{ and } P(\mathcal{S}'_0 , \mathcal{S}'_e) \implies \exists U \in \mathcal{U}(\H): (\mathcal{S}'_0 , \mathcal{S}'_e) = U \cdot (\mathcal{S}_0 , \mathcal{S}_e).
\end{equation}
\end{definition}

The idea of this definition is to find a property that selects a single orbit in $\mathcal{K}_0 \times \mathcal{K}_e$, hence a unique physical theory from the point of view of the predictions. We need to require $\mathcal{K}_0$ and $\mathcal{K}_e$ to be determined kinds, because otherwise it is not even possible in principle to relate $\mathcal{S}_0$ to $\mathcal{S}'_0$, or $\mathcal{S}_e$ to $\mathcal{S}'_e$, \textit{via} a unitary transformation.

An instance of such a determining property $P$ would be the specification of values for a complete set of invariants\footnote{In the context of quantum mechanics, the notion of complete set of invariants has mainly been used so far to characterize the entanglement classes of a composite system. In this case, the group is $\mathcal{U}(\H) \times \dots \times \mathcal{U}(\H)$ acting on $\H_1 \otimes \dots \otimes \H_n$ by local unitaries. Each orbit of this action constitute a possible class of entanglement for the system. We know that there always exists a finite set of homogeneous polynomials in the coefficients of the state that form a complete set of invariants, in virtue of Hilbert's basis theorem. These polynomials are even explicitly known for small dimensions or few subsystems~\cite[Chapter 17.4]{bengtsson2017geometry}\cite{makhlin2002nonlocal, zhou2012local, dobes2025local}.}, a central concept of invariant theory~\cite{mumford1994geometric}. This is, in fact, what was done in the example~\ref{kind_HSF}. Retrospectively, the use of invariants to discriminate between different orbits was also at the heart of the proof of proposition \ref{infinity}. Equipped with this new notion, we are now able to rephrase Cotler \textit{et al.}'s theorem in an elegant way. 

\begin{example}
The property of $K$-locality determines $\mathcal{K}_{\text{Herm}(\sigma)} \times \mathcal{K}_{\text{TPS}(n; d_1, \dots, d_n)}$ according to Cotler \textit{et al.}'s theorem, under the assumptions of the proof.
\end{example} 

Although $\mathcal{K}_0$ and $\mathcal{K}_e$ are determined kinds, $\mathcal{K}_0 \times \mathcal{K}_e$ may contain numerous orbits, as many as there are ways for $\mathcal{S}_0$ and $\mathcal{S}_e$ to ‘stand’ with respect to each other. To picture all these possible mutual configurations, it may be convenient to systematically rotate the pair so that the input structure is mapped to a fixed $\mathcal{S}_0$, and then simply study where $\mathcal{S}_e$ lands. This is what Cotler \textit{et al.} did when introducing the notion of ‘TPS equivalent with respect to a fixed $\hat{H}$’ (second part of definition \ref{equivalence}) and this also grounds the intuition for Stoica's method. The proposition below allows us to move from one perspective to the other. In what follows, $\stab(\mathcal{S}_0)$ denotes, as in section \ref{stoica}, the symmetries of the structure $\mathcal{S}_0$, namely the set $\{ U \in \mathcal{U}(\H) \mid U \cdot \mathcal{S}_0 = \mathcal{S}_0 \}$ and $[\mathcal{S}_e]_{\stab(\mathcal{S}_0)}$ denotes the orbit of $\mathcal{S}_e \in \mathcal{K}_e$ under the action of $\stab(\mathcal{S}_0)$ on $\mathcal{K}_e$.

\begin{proposition}[From $3^\text{rd}$ to $1^\text{st}$ person perspective] \label{3rd to 1st}
Let $\mathcal{K}_0$ and $\mathcal{K}_e$ be two determined kinds, $P$ a unitary-invariant property on $\mathcal{K}_0 \times \mathcal{K}_e$ and $\mathcal{S}_0$ a fixed $\mathcal{K}_0$-structure. There exists a bijection $\pi_{\mathcal{S}_0}$ between the set of orbits in $\mathcal{K}_0 \times \mathcal{K}_e$ under the action of $\mathcal{U}(\H)$ and the set of orbits in $\mathcal{K}_e$ under the action of $\stab(\mathcal{S}_0)$. Moreover, this map is compatible with $P$. Said differently, we have the following commutative diagram:
\begin{equation}
\begin{tikzcd}
\faktor{\mathcal{K}_0 \times \mathcal{K}_e}{\mathcal{U}(\H)} \arrow[rd, "P(\cdot {,} \cdot)"'] \arrow[r, "\pi_{\mathcal{S}_0}", "\sim"'] & \faktor{\mathcal{K}_e}{\stab(\mathcal{S}_0)} \arrow[d, "P(\mathcal{S}_0 {,} \cdot)"] \\
& \{ \mathrm{True}, \mathrm{False} \}
\end{tikzcd}
\end{equation}
Consequently, $P$ determines the product kind $\mathcal{K}_0 \times \mathcal{K}_e$ if and only if the set $\{ \mathcal{S}_e \mid P(\mathcal{S}_0, \mathcal{S}_e) \}$ is the orbit of a single element under the action of $\stab(\mathcal{S}_0)$ on $\mathcal{K}_e$. 
\end{proposition}

\begin{proof}
Let's first build the map $\pi_{\mathcal{S}_0}$. For all $\mathcal{S}'_0 \in \mathcal{K}_0$, there exists a unitary $U_{\mathcal{S}'_0, \mathcal{S}_0}$ mapping $\mathcal{S}'_0$ to $\mathcal{S}_0$, because $\mathcal{K}_0$ is determined. We define:
\begin{equation}
    \begin{array}{lcll} \Pi_{\mathcal{S}_0} : & \mathcal{K}_0 \times \mathcal{K}_e &  \longrightarrow &  \faktor{\mathcal{K}_e}{\stab(\mathcal{S}_0)} \\ \\
    & (\mathcal{S}'_0, \mathcal{S}'_e) & \longmapsto  & [U_{\mathcal{S}'_0, \mathcal{S}_0} \cdot \mathcal{S}'_e]_{\stab(\mathcal{S}_0)}.  \end{array}
\end{equation} 
This definition is consistent, namely it does not depend on the particular choice of $U_{\mathcal{S}'_0, \mathcal{S}_0}$, because another unitary $V_{\mathcal{S}'_0, \mathcal{S}_0}$ mapping $\mathcal{S}'_0$ to $\mathcal{S}_0$ is related to $U_{\mathcal{S}'_0, \mathcal{S}_0}$ by an element of $\stab(\mathcal{S}_0)$ (as $U_{\mathcal{S}'_0, \mathcal{S}_0} V^\dagger_{\mathcal{S}'_0, \mathcal{S}_0}$ fixes $\mathcal{S}_0$).

The map $\Pi_{\mathcal{S}_0}$ is surjective. Indeed, for all $\mathcal{S}'_e \in \mathcal{K}_e$, 
\begin{equation}
\Pi_{\mathcal{S}_0}(\mathcal{S}_0, \mathcal{S}'_e) = [U_{\mathcal{S}_0, \mathcal{S}_0} \cdot \mathcal{S}'_e]_{\stab(\mathcal{S}_0)} = [\mathcal{S}'_e]_{\stab(\mathcal{S}_0)},    
\end{equation}
since $U_{\mathcal{S}_0, \mathcal{S}_0} \in \stab(\mathcal{S}_0)$. Moreover, $\Pi_{\mathcal{S}_0}$ is unitary-invariant, \textit{i.e.} it is constant on the orbits in $\mathcal{K}_0 \times \mathcal{K}_e$ under the action of $\mathcal{U}(\H)$. To see this, let $U \in \mathcal{U}(\H)$ and observe that $U_{\mathcal{S}'_0, \mathcal{S}_0} U^\dagger$ is a unitary mapping $U \cdot \mathcal{S}'_0$ to $\mathcal{S}_0$. Hence:
\begin{equation}
    \Pi_{\mathcal{S}_0}(U \cdot \mathcal{S}'_0 , U \cdot \mathcal{S}'_e) = [U_{\mathcal{S}'_0, \mathcal{S}_0} U^\dagger \cdot U \cdot \mathcal{S}'_e]_{\stab(\mathcal{S}_0)} = [U_{\mathcal{S}'_0, \mathcal{S}_0} \cdot \mathcal{S}'_e]_{\stab(\mathcal{S}_0)} = \Pi_{\mathcal{S}_0}(\mathcal{S}'_0 ,\mathcal{S}'_e).
\end{equation}
Furthermore, $\Pi_{\mathcal{S}_0}$ takes a different value on each orbit, because:
\begin{align}
    \Pi_{\mathcal{S}_0}(\mathcal{S}'_0, \mathcal{S}'_e) = (\mathcal{S}''_0, \mathcal{S}''_e) 
    \Rightarrow & \quad [U_{\mathcal{S}'_0, \mathcal{S}_0} \cdot \mathcal{S}'_e]_{\stab(\mathcal{S}_0)} = [U_{\mathcal{S}''_0, \mathcal{S}_0} \cdot \mathcal{S}''_e]_{\stab(\mathcal{S}_0)} \nonumber \\
    \Rightarrow & \quad \exists V \in \stab(\mathcal{S}_0) : \mathcal{S}'_e = U^\dagger_{\mathcal{S}'_0, \mathcal{S}_0} V U_{\mathcal{S}''_0, \mathcal{S}_0} \cdot \mathcal{S}''_e \\
    \Rightarrow & \quad \exists U \in \mathcal{U}(\H) : (\mathcal{S}'_0 , \mathcal{S}'_e) = U \cdot (\mathcal{S}''_0 , \mathcal{S}''_e) \nonumber
\end{align}
As a consequence, $\Pi_{\mathcal{S}_0}$ can be factorized into a bijective map 
\begin{equation}
    \pi_{\mathcal{S}_0} : \faktor{\mathcal{K}_0 \times \mathcal{K}_e}{\mathcal{U}(\H)} \overset{\sim}{\longrightarrow} \faktor{\mathcal{K}_e}{\stab(\mathcal{S}_0)}.
\end{equation}

It remains to show that the diagram commutes. First, note that, by unitary-invariance, $P$ is well-defined on $\faktor{\mathcal{K}_0 \times \mathcal{K}_e}{\mathcal{U}(\H)}$. For the same reason,
\begin{equation}
    P(\mathcal{S}_0, \cdot) \; : \; \mathcal{S}_e \longmapsto P(\mathcal{S}_0, \mathcal{S}_e)
\end{equation}
is well-defined on $\faktor{\mathcal{K}_e}{\stab(\mathcal{S}_0)}$. Finally, 
\begin{equation}
    P\left(\mathcal{S}'_0 , \mathcal{S}'_e \right) 
    =  P\left(U_{\mathcal{S}'_0, \mathcal{S}_0} \cdot \mathcal{S}'_0 \; , \; U_{\mathcal{S}'_0 , \mathcal{S}_0} \cdot \mathcal{S}'_e \right) 
    = P(\mathcal{S}_0, U_{\mathcal{S}'_0, \mathcal{S}_0} \cdot \mathcal{S}'_e ) 
    = P(\mathcal{S}_0, \pi_{\mathcal{S}_0}( \mathcal{S}'_0 , \mathcal{S}'_e )).
\end{equation}
\end{proof}

\begin{example}
Because $\mathcal{K}_{\text{Herm}(\sigma)}$ and $\mathcal{K}_{\text{TPS}(n; d_1, \dots, d_n)}$ are determined kinds, proving that $K$-locality determines the product $\mathcal{K}_{\text{Herm}(\sigma)} \times \mathcal{K}_{\text{TPS}(n; d_1, \dots, d_n)}$ amounts to showing that, for a fixed $\hat{H}$, all the TPSs that make $\hat{H}$ $K$-local are related by a symmetry of $\hat{H}$, thereby justifying Cotler \textit{et al.}'s claim in~\cite{cotler2019locality} (recall footnote \ref{twofold_definition}).
\end{example}

Intuitively, this proposition can be understood as follows. A $\mathcal{K}_0$-structure will in general not be a perfect ‘reference frame’ in $\H$: from the perspective of $\mathcal{S}_0 \in\mathcal{K}_0$, other surrounding structures are distinguished only up to $\stab(\mathcal{S}_0)$, thus $\mathcal{S}_0$ ‘sees’ a coarse-grained Hilbert space with an effective symmetry group $\mathcal{U}(\H) / \stab(\mathcal{S}_0)$. Said differently, characterizing up to $\stab(\mathcal{S}_0)$ a preferred structure $\mathcal{S}_e$ through its relations to a fixed $\mathcal{S}_0$ is the best thing one can hope for. However, if the unitary group happens to have a free action on $\mathcal{K}_0$, meaning that $\stab(\mathcal{S}_0) =\{ \bbone \}$, then the $\mathcal{K}_0$-structures are sufficiently rich to discriminate between any two distinct structures of another kind. In this case, the absolute perspective mentioned in \ref{uniqueness} turns out to be justified, according to the following observation.

\begin{corollary} \label{relational_to_absolute}
If $\mathcal{U}(\H)$ acts freely on $\mathcal{K}_0$, a property $P$ determines the product kind $\mathcal{K}_0 \times \mathcal{K}_e$ if and only if for all $\mathcal{S}_0 \in \mathcal{K}_0$, there exists a unique $\mathcal{S}_e \in \mathcal{K}_e$ such that $P(\mathcal{S}_0, \mathcal{S}_e)$ holds.
\end{corollary}

In passing, note that the existence of the bijection $\pi_{\mathcal{S}_0}$ solves the problem of knowing when the product of two kinds is determined.
\begin{corollary}
    The product kind $\mathcal{K}_0 \times \mathcal{K}_e$ is determined (without imposing any additional $P$) if and only if $\stab(\mathcal{S}_0)$ acts transitively on $\mathcal{K}_e$.
\end{corollary}

\subsection{Can $(\hat{H}, \ket{\psi})$ uniquely determine a TPS?}\label{sec:maintheorem}

Corollary \ref{relational_to_absolute} is important in the context of HSF, where the input structures are of the kind $\smash{\mathcal{K}_{\text{HSF}(\sigma,\Lambda)}}$, as we have shown in example \ref{kind_HSF} that the unitary group acts freely on this determined kind. Thus, finding a property that selects a unique triplet $(\hat{H}, \ket{\psi}, \mathcal{S}_e)$ up to unitary equivalence---where $\mathcal{S}_e$ can be of any kind, typically a TPS---boils down to fixing some $\smash{(\hat{H}, \ket{\psi}) \in \mathcal{K}_{\text{HSF}(\sigma,\Lambda)}}$ and finding a property $P(\hat{H}, \ket{\psi}, \mathcal{S}_e)$ holding for exactly one $\mathcal{S}_e$. It is in this sense that Stoica's absolute attitude towards structures can become justified. 

However, we contest one of Stoica's claims, according to which any emergent structure $\mathcal{S}_e$ characterized by a unitary-invariant property of $(\hat{H}, \ket{\psi}, \mathcal{S}_e)$ is not unique (in the absolute sense), provided that it is physically relevant (a notion introduced in~\cite[Theorem 2]{stoica20213d}, meaning in short that $\mathcal{S}_e$ distinguishes different states $\ket{\psi} \neq \ket{\psi'}$). Obviously, Stoica's method generates an abundance of triples $(\hat{H}, \ket{\psi}, \mathcal{S}_e)$ that are all unitary-equivalent because they are obtained from a first triple by applying global unitaries, so there is no chance to show any non-uniqueness in the relational sense. But, even in the absolute sense, the argument is flawed. By applying a symmetry of $\hat{H}$ to $(\hat{H}, \ket{\psi}, \mathcal{S}_e)$, one obtains indeed a new triple $(\hat{H}, U\ket{\psi}, U\cdot \mathcal{S}_e)$, but now the input structure has been modified, thus it does not prove that a \emph{fixed} $(\hat{H}, \ket{\psi})$ can not determine uniquely $\mathcal{S}_e$ in the sense of corollary \ref{relational_to_absolute}. In short, his approach does not treat $\hat{H}$ and $\ket{\psi}$ on an equal footing: the former is considered in the absolute sense, but the latter in the relational sense.

As a concrete example, in the case of the TPS, Stoica's method allows one to construct many TPSs that all satisfy some set of unitary-invariant conditions with respect to $\hat{H}$, but nothing guarantees that these TPSs are not distinguished by $\ket{\psi}$. In fact, as we shall argue now, a pair $(\hat{H}, \ket{\psi})$ \emph{is} enough to select a unique TPS $\mathcal{T}$ up to unitary equivalence on $(\hat{H}, \ket{\psi}, \mathcal{T})$.

\begin{theorem}\label{main_theorem}
There exists a unitary-invariant property $P$ that determines the product kind 
$\mathcal{K}_{\mathrm{HSF}(\sigma,\Lambda)} 
\times 
\mathcal{K}_{\mathrm{TPS}}(n; d_1,\dots,d_n)$, if the spectrum $\sigma$ is non-degenerate and all projections $\lambda_i \in \Lambda$ are non-zero.
\end{theorem}

\begin{proof}
According to corollary \ref{relational_to_absolute}, we just need to find a unitary-invariant property $P$ such that, for \emph{fixed} $\hat{H}$ and $\ket{\psi}$, there exists a unique TPS $\mathcal{T}$ such that $P(\hat{H}, \ket{\psi}, \mathcal{T})$ holds. Let then $(\hat{H}, \ket{\psi}) \in \mathcal{K}_{\mathrm{HSF}(\sigma,\Lambda)}$. Under the assumption made on $\sigma$ and $\Lambda$, we have: 
\begin{equation} \label{polynomials_generate_H}
\{ R(\hat{H}) \ket{\psi} \mid R\in \mathbb{C}[X] \} = \mathcal{H},
\end{equation}
where $\mathbb{C}[X]$ is the space of polynomials in one variable with complex coefficients. Indeed, diagonalizing $\hat{H} = \sum_k \omega_k \ket{\omega_k}\!\!\bra{\omega_k}$ yields $R(\hat{H}) = \sum_k R(\omega_k) \ket{\omega_k}\!\!\bra{\omega_k}$, which can reach any operator $\sum_k a_k \ket{\omega_k}\!\!\bra{\omega_k}$ diagonal in $(\ket{\omega_k})_k$ by choosing the appropriate interpolating polynomial $R$ mapping any $\omega_k$ to $a_k$ (this is possible since $\hat{H}$'s spectrum is non-degenerate). Moreover, by assumption, $\ket{\psi}$ has full support on $\hat{H}$'s eigenbasis $(\ket{\omega_k})_k$, meaning that for all $k$, $c_k \equiv \braket{\psi \vert \omega_k} \neq 0$. For this reason 
\begin{equation}
R(\hat{H}) \ket{\psi} = \left( \sum_l a_l \ket{\omega_l}\!\!\bra{\omega_l} \right)  \sum_k c_k \ket{\omega_k} = \sum_k a_k c_k \ket{\omega_k} 
\end{equation}
can reach any vector in $\mathcal{H}$. 

If $L=(s_{R,i})_{R \in \mathbb{C}[X], \;  1\leq i \leq n}$ be a collection of reals numbers in $[0,\log d_i]$ indexed by $\mathbb{C}[X] \times \{1, \dots, n \}$, we define $P_L$ to be the following unitary-invariant property:
\begin{equation}
P_L(\hat{H}, \ket{\psi}, \mathcal{T}) \quad : \quad \forall R \in \mathbb{C}[X], \; \forall i \in \{1, \dots, n\}, \quad S \left[ \rho_i^{\mathcal{T}}\left( R(\hat{H})\ket{\psi} \right) \right] = s_{R,i},
\end{equation}
where $S$ denotes the von Neumann entropy and $\rho_i^{\mathcal{T}}(\ket{\Phi}) = \tr_{\bar{i}} \left( \ket{\Phi}\!\!\bra{\Phi} \right)$ the $i^\text{th}$ partial state of $\ket{\Phi}$ in the TPS $\mathcal{T}$. Intuitively, $P_L$ amounts to imposing the entanglement entropies of all $R(\hat{H})\ket{\psi}$ with respect to the TPS $\mathcal{T}$. 

Now, let $\mathcal{T}, \mathcal{T}' \in \mathcal{K}_{\mathrm{TPS}}(n; d_1,\dots,d_n)$ and suppose that $P_L(\ket{\psi}, \hat{H}, \mathcal{T})$ and $P_L(\ket{\psi}, \hat{H}, \mathcal{T}')$ hold. Let $U$ be a unitary such that $U \cdot \mathcal{T} = \mathcal{T}'$. Such a $U$ exists because $\mathcal{T}$ and $\mathcal{T}'$ can be entirely characterized by specifying two properly labeled orthonormal bases of $\mathcal{H}$, so it suffices to choose a $U$ mapping one basis to the other. Let's show that $U=U_1 \otimes \dots \otimes U_n$, which would imply $\mathcal{T} = \mathcal{T}'$.

Using \eqref{polynomials_generate_H}, we see that the entanglement entropies of \emph{any} vector in $\H$ must have the same values in $\mathcal{T}$ and $ \mathcal{T}'$, if $P_L$ holds for both TPSs.
Equivalently, this means that for all $\ket{\Phi} \in \mathcal{H}$, $\ket{\Phi}$ and $U^\dagger \ket{\Phi}$ have the same entanglement entropies with respect to $\mathcal{T}$. In particular, $U$ must map pure tensors in $\mathcal{T}$ to pure tensors in $\mathcal{T}$. Indeed, being a pure tensor is characterized by all partial states having $0$ von Neumann entropy. Since any operator mapping pure tensors to pure tensors is a product of single-site operators (see lemma~\ref{lemma_tensor} in the appendix), we conclude that $U$ is a product of single-site unitaries.
\end{proof}

Just like for the proof of proposition~\ref{infinity}, we could equally have used the purities, or even the full list of eigenvalues of the $\rho_i$ instead of the von Neumann entropies, but we chose the latter because it is easier to interpret physically. Nevertheless, the property $P_L$ has admittedly less immediate physical meaning than the $K$-locality requirement, for instance. The interest of this theorem is first and foremost to give a proof of principle that $(\hat{H}, \ket{\psi})$ is enough structure to uniquely determine $\mathcal{T}$; it paves the way for a search for more meaningful properties.

Clearly, not every $L$ gives rise to a satisfiable $P_L$, in the sense that $P_L$ may not correspond to the entanglement entropies of any state $\ket{\psi}$. Likewise, not every $N\times N$ matrix is a Gram matrix that can be used to select a unique orbit among the families of $N$ vectors in $\mathcal{H}$. Determining the set of collections $L$ that really encode the entanglement properties of some TPS is a special case of a famously hard problem known as the ‘quantum marginal problem’~\cite{schilling2014quantum, tyc2015quantum, yu2021complete}. Fortunately, our argument does not need such a characterization: the situation is similar to Cotler \textit{et al.}’s context, in which it suffices to show that, \emph{if} there is a TPS making $\hat{H}$ $K$-local, then this TPS is unique.

Interestingly, the proof still holds if one merely imposes the entanglement entropies of the $R(\hat{H})\ket{\psi}$ for $R \in (\mathbb{Q}+i\mathbb{Q})[X]$ (\textit{i.e.} complex polynomials with coefficients having rational real and imaginary parts). By density of $(\mathbb{Q}+i\mathbb{Q})[X]$ in $\mathbb{C}[X]$ and by continuity of the von Neumann entropy, this is sufficient to impose the entanglement entropies of all vectors in $\H$, from which the theorem follows. It means that we only need to require a countable set of conditions to uniquely determine $\mathcal{T}$ out of $\hat{H}$ and $\ket{\psi}$. Said differently, we have found a countable complete set of invariants for the product kind $\mathcal{K}_{\mathrm{HSF}(\sigma,\Lambda)} 
\times 
\mathcal{K}_{\mathrm{TPS}}(n; d_1,\dots,d_n)$.

Is it possible to relax the assumption on $\sigma$ and $\Lambda$? Not entirely, for sure. If $\hat{H}$'s spectrum is too degenerate, or if $\ket{\psi}$ has too narrow support on $\hat{H}$'s eigenbasis, then the set of vectors $\{ R(\hat{H}) \ket{\psi} \mid R\in \mathbb{C}[X] \}$ will be too small to grasp $\mathcal{T}$. Think, for instance, of the extreme situations $\hat{H}= \bbone$ or $\ket{\psi} = \ket{\omega_i}$: in both cases, only a 1-dimensional vector space is spanned. However, the question of finding the minimal set of vectors able to determine a TPS is open.

\section{Conclusion}

The ability to uniquely select a theory in virtue of a physically motivated principle has proved to be a fruitful and satisfactory approach in theoretical physics. Einstein’s tensor is the only tensor at most second order in the metric that can consistently be equated to the stress-energy tensor \cite{lovelock1971einstein}; the quantum fields are the only possible fields of operators that transform correctly under the Poincaré group; Schrödinger’s equation is the only possible Markovian unitary time evolution in a Hilbert space.
The various reconstructions of Hilbert-space quantum theory from operational principles~\cite{hardy2001quantum,dakic2009quantum,chiribella2011informational,masanes2011derivation,hardy2013reconstructing,hohn2017quantum,hohn2017toolbox,jia2018quantum} and the singling-out of bosonic and fermionic statistics over other exchange statistics (under the requirement of complete invariance or invariance under quantum permutations~\cite{mekonnen2025invariance}) offer additional recent quantum information theoretical examples.

What does it mean, though, for a structure to be uniquely constrained by a property? This question is particularly pressing in the context of quantum mechanics, from concerns in decoherence theory and quantum mereology, to Hilbert space fundamentalism and the quest for a background-independent quantum theory. In this work, we have first focused on the specific problem of whether a Hamiltonian can uniquely select a preferred tensor product structure, by reviewing and clarifying two seemingly contradictory theorems by Cotler \textit{et al.} and Stoica in sections \ref{cotler} and \ref{stoica}. We have then broadened the discussion in section~\ref{characterizing}, where our main contributions were: (i) to dissolve the apparent tension between the two results; (ii) to argue that the appropriate notion of uniqueness in physics should be understood relationally (\textit{i.e.} in the case of quantum mechanics, up to a global unitary transformation on the full set of structures considered) because two theories yielding the same predictions should be deemed equivalent; (iii) to show that the question boils down to finding unitary-invariant properties that single out an orbit in a Cartesian product of structures under the unitary group; (iv) to reduce the problem of uniqueness for a pair of structures to a problem for a single structure with respect of another fixed one (proposition \ref{3rd to 1st}); (v) to explain why both Cotler \textit{et al.} and Stoica's arguments are not sufficient to solve the specific question of whether $(\hat H, \ket \psi)$ can select a preferred TPS $\mathcal{T}$; (iv) most importantly, to prove that $(\hat H, \ket \psi)$ is enough structure to single out a TPS (Theorem~\ref{main_theorem}). The latter revives the hope that HSF is tenable, and, in particular, it possibly justifies the approaches to spacetime emergence that rely on a preferred TPS from which spatiotemporal structures are obtained.

Is there a general recipe to tell, given two kinds of structures, whether their relations allow for the existence of a property that uniquely characterizes a pair of them, up to unitary equivalence? Or do we have to work on a case by case basis, because the situation fundamentally depends on nature of the structure involved? \textit{A priori}, it seems that proving that a given property works is a hard task in general; see for example how tough is Cotler \textit{et al.}'s proof in the case of a pair $(\hat H, \mathcal T)$! It would also be interesting to search for situations in which there exists no complete set of invariants, meaning that no property can possibly select a unique orbit in the Cartesian product of structures.

An important limitation remains. Like Cotler \textit{et al.}’s and Stoica’s, our construction is currently restricted to finite-dimensional Hilbert spaces and assumes prior knowledge of the number $n$ of tensor factors as well as their dimensions $(d_i)_{1 \leq i \leq n}$. When aiming to explain the emergence of subsystems in quantum theory, it is certainly undesirable to have to specify in advance their dimensionality. Relaxing this assumption, and applying the formalism to other kinds of structures, could turn out to have wide applications, in particular in the fields of quantum mereology and quantum gravity. 

The relational philosophy of section~\ref{characterizing} also clearly resonates with the quantum reference frames (QRF) approach \cite{bartlett_rudolph_spekkens_2007, loveridge_miyadera_busch_2018, giacomini_castro-ruiz_brukner_2019, vanrietvelde_hoehn_giacomini_castro-ruiz_2020, de_la_hamette_galley_2020, ali2022quantum, carette_glowacki_loveridge_2023, lake_miller_2023, castro2025relative}. In particular, the map $\pi_{\mathcal{S}_0}$ introduced in proposition~\ref{3rd to 1st} temptingly reminds how one moves from the perspective-neutral state to the state in the reference frame of a system, namely by fixing the value of some observable attached to this system \textit{via} a ‘reduction map’~\cite{vanrietvelde_hoehn_giacomini_castro-ruiz_2020, de2021perspective, ali2022quantum}. It would be interesting to see whether this result can be put to fruitful use in the field as a whole.

The scope of the present work might even extend beyond this. The methodology of section \ref{characterizing} was based on geometrical arguments. In Félix Klein's approach to geometry—a view he pioneered as soon as 1872 \cite{klein1893vergleichende} to unify the different geometries existing at his time (the Erlangen program) and that has remained highly influential up to now—a geometry \emph{is} exactly a set $E$ equipped with a group $G$ acting on it. For the purposes of this paper, we have restricted ourselves to the case $(E,G) = (\H, \mathcal{U}(\H))$, although the results of section \ref{insights} would remain valid for any other geometrical space. Considering that our four most fundamental theories in contemporary physics are all expressed in the language of geometry\footnote{Special relativity and general relativity are formulated in the framework of differential (Lorentzian) geometry, quantum mechanics in terms of the geometry of complex Hilbert spaces and its canonical extension to the tensor product structure, and quantum field theory in terms of fibre bundles of operators over a Lorentzian, possibly curved manifold.}, these ideas could in principle find applications far beyond quantum physics, in particular when setting $(E,G) = (M, \mathrm{Diff}(M))$.

\section*{Acknowledgements}

The authors thank \v{C}aslav Brukner for several discussions about these ideas. AS would like to thank Daniel Ranard and Cristi Stoica for an unexpected email exchange that dramatically enhanced the relevance of this work, as well as the CPT Marseille---especially Thierry Masson and Serge Lazzarini---for hosting a fantastic 1-month visit during which part of this work was conducted. Special thanks go to Béranger Seguin for illuminating discussions on the foundations of geometry and category theory, Nasra Daher Ahmed for asking a key question that unlocked many doors, and Simon Fuchs for careful re-reading and sharp remarks. GF and ADB thank Ofek Bengyat and Carlo Cepollaro for discussions on this topic at the \href{https://www.iqoqi-vienna.at/kefalonia-foundations/participants-kefalonia-2023}{Kefalonia Foundations 2023} workshop. GF also thanks Tales Rick Perche for a careful reading of the draft. ADB finally thanks Emil Broukal for getting him interested in this question again.

AS's research is funded by the Austrian Science Fund (FWF) through BeyondC grant [10.55776/ F71]. GF is thankful for financial support from the Olle Engkvist Foundation (no.225-0062) and acknowledges support by a STSM Grant from COST Action CA23130. ADB's research was funded within the QuantERA II Programme that has received funding
from the European Union’s Horizon 2020 research and innovation programme under Grant
Agreement No 101017733, and from the Austrian Science Fund (FWF), projects I-6004 and
ESP2889224 as well as Grant No. I 5384.

\addtocontents{toc}{\protect\setcounter{tocdepth}{0}}

\appendix
\section{Proof of lemma~\ref{lemma_tensor}}
\label{appendix_tensor}

In a tensor product space, a pure tensor is an element of the form $\ket{x_1} \otimes \dots \otimes \ket{x_n}$ (we call it a product state if moreover it is normalized).

\begin{lemma}\label{lemma_tensor}
An operator $A \in \mathcal{L}(\mathcal{H}_1 \otimes \dots \otimes \mathcal{H}_n)$ is a product of single-site operators $A_1 \otimes \dots \otimes A_n$ if and only if $A$ maps pure tensors to pure tensors.
\end{lemma}

\begin{proof}
The direct implication is immediate. To show the converse, let $(\ket{e_{k_i}^{(i)}})_{1\leq k_i \leq d_i}$ be an orthonormal basis of $\mathcal{H}_i$ for all $i \in \{1, \dots, n\}$. A pure tensor is a vector of the form:
    \begin{equation}
        \ket x = \left(\sum_{k_1} x^{(1)}_{k_1} \ket{e_{k_1}^{(1)}} \right) \otimes \dots \otimes \left(\sum_{k_n} x^{(n)}_{k_n} \ket{e_{k_n}^{(n)}} \right) = \sum_{k_1, \dots , k_n} x^{(1)}_{k_1} \dots x^{(n)}_{k_n} \ket{e_{k_1}^{(1)} \dots e_{k_n}^{(n)} }, 
    \end{equation}
for some families of coefficients $\big(x^{(1)}_{k_1} \big)_{1 \leq k_1 \leq d_1}, \dots, \big(x^{(n)}_{k_n} \big)_{1 \leq k_n \leq d_n}$. A generic operator in $\mathcal{L}(\mathcal{H}_1 \otimes \dots \otimes \mathcal{H}_n)$ can be written as:
    \begin{equation}
        A = \sum_{\substack{k_1, \dots , k_n \\ k'_1, \dots , k'_n}} A_{\substack{k_1, \dots , k_n \\ k'_1, \dots , k'_n}} \ket{e_{k_1}^{(1)} \dots e_{k_n}^{(n)} }\!\!\bra{e_{k'_1}^{(1)} \dots e_{k'_n}^{(n)} },
    \end{equation}
which is of the form $A_1 \otimes \dots \otimes A_n$ if and only if there exist some matrices $(a_{k_i, k'_i}^{(i)})_{k_i, k'_i}$ for each $i$ such that:
    \begin{equation}
        A_{\substack{k_1, \dots , k_n \\ k'_1, \dots , k'_n}} = a_{k_1, k'_1}^{(1)} \dots a_{k_n, k'_n}^{(n)}.
    \end{equation}

Now, if an operator $A$ in $\mathcal{L}(\mathcal{H}_1 \otimes \dots \otimes \mathcal{H}_n)$ maps pure tensors to pure tensors, then in particular
 \begin{equation}
        A \ket{e_{k'_1}^{(1)} \dots e_{k'_n}^{(n)}} = \sum_{k_1, \dots , k_n } A_{\substack{k_1, \dots , k_n \\ k'_1, \dots , k'_n}} \ket{e_{k_1}^{(1)} \dots e_{k_n}^{(n)} }
    \end{equation}
must be a pure tensor. Consequently, for all $k'_1, \dots, k'_n$, there exist some families of coefficients $\big(x^{(1)}_{k_1, k'_1} \big)_{1 \leq k_1 \leq d_1}, \dots, \big(x^{(n)}_{k_n, k'_n} \big)_{1 \leq k_n \leq d_n}$ such that:
    \begin{equation}
        A_{\substack{k_1, \dots , k_n \\ k'_1, \dots , k'_n}} = x^{(1)}_{k_1, k'_1} \dots x^{(n)}_{k_n, k'_n}.
    \end{equation}
This proves that $A$ is indeed a product of local operators.

\end{proof}
\printbibliography
\end{document}